# AS-PUMA : ANYCAST SEMANTICS IN PARKING USING METAHEURISTIC APPROACH

## Rahul K Dixit[1] and Rahul Johari[2]

[1] Research Scholar, USICT, GGS Indraprastha University, Delhi-110078
rahul.1234dixit@gmail.com
[2] Assistant Professor, USICT, GGS Indraprastha University, Delhi-110078
rahuljohari@hotmail.com

## ABSTRACT

*The number of vehicle used in the world are increasing day by day resulting in the obvious problem of parking of these vehicle's in residential and vocational areas. We perceive the problem of vehicles parking in vocational establishments / malls. Today majority of parking systems are manual parking systems where in, on the spot, parking of the vehicle is done and a parking slip is generated and handed over to customer. This is cumbersome technique wherein various parking attendants in the parking areas manually keeps on informing the Parking inspector on how many free parking slots available so that only that many number of parking slips/tickets are generated as the number of free parking slots. We address the problem of parking in Delay Tolerant Network (DTN) by proposing metaheuristic driven approach of Ant Colony optimization (ACO) technique with anycast semantics models . Here we propose the parking architecture to solve the problem of parking especially in commercial areas with their design diagrams . In this architecture we apply the delivery model to deliver the packet correctly to the intended receiver. Using this we can book various parking's through remote areas so that the customer can get the information about availability of various parking's inside an area and the parking fare for each category of the automobile. Using this architecture the customer can get the prior knowledge about various vacant parking slots inside a parking area and he can book the corresponding parking from his location.*

## 1. INTRODUCTION

The MANETs (Mobile Ad hoc Networks) [1] address challenges related to mobility and low battery life when there is no pre-existing communication infrastructure. In most of these routing techniques there is an underlying assumption that there always exists a connected path from source to destination. Delay/Disruption Tolerant Networking (DTN) [2], variant of MANET adopts the concept of intermittent networks that may suffer frequent disconnections. A Delay-Tolerant Network is those networks that may experience frequent disruptions, long- duration partitioning and may never have an end-to-end connected path.[4,5,6]. In this work, we deal with the problem of routing in a Delay Tolerant Network. In contrast to the traditional routing techniques of MANET's where the aim is fast delivery of a message, here the aim is the delivery itself. Since the links are available for a time being so that packet must be stored and forwarded afterwards, delays are inevitable in such networks. Therefore, the design of protocols for those networks becomes a unique challenge. Routing in DTNs is one of the key components and remains open for discussion.

There are two evolutionary approaches we can solve our problem (Routing in DTN) one is by genetic Algorithms and Second is Ant colony Optimization

## 1.1 GENETIC ALGORITHMS:

GA are global search and optimization Techniques modelled from natural selection, genetic and evolution. The GA implements this process through coding and special operators approach. The basic rules of GA were first observed [9]. Excellent reference on GAs and their applications were found [10]. A genetic algorithm maintains the population of candidate solutions, where each candidate solution is usually coded as binary string called a chromosome. The best choice of coding has been shown to be a binary coding [9]. The chromosomes set forms a population, which is estimated and ranked by few particular function called as fitness estimation function. The fitness estimation function play a critical role in GAs because it provides information about how good each candidate is. The initial population is usually originated randomly. The evolution from one generation to the next one involves mainly three steps: fitness estimation, selection and reproduction [11].

## 1.2: ACO[3],[7]

Aco is an algorithm founded on the experience of real ants in searching a shortest path from their nest to food. The algorithm exploits the experience of real ants when they search for food. It has been seen that the ants laying down Pheromone at the time of searching the food. Once when the food is discovered by the ants again they deposit the pheromone in its path while returning to its nest. By doing this, ants following the shorter path are expected to return earlier and hence the pheromone level becomes low where, the following the longer distance. ACO takes the infusion from the real ants. These ants drop down the pheromone on the path to get some path which is compatible to followed colony member also. However, the pheromone evaporation is done at a constant rate after a certain interval of time that's why only those path are remains (kept marked) which was most frequently visited and rarely visited path are lost because of the lack of pheromone deposit on that path hence as a result the other ants are intended to follow repeatedly used paths only. Thereby, all the ants in a search of food starts their journey can get the information provided by the previously visitor ants and are monitored to follow the shortest path as indicated by the pheromone deposit by other ants

We propose a routing approach using Ant Colony Optimization (ACO) [3, 7], wherein the route and destination decision influence many parameters simultaneously. As ACO suggests that in the natural world, at first the ants move randomly, and upon finding food return to their colony while drop down pheromone footprints, same way our problem of parking vehicles proceeds. But here we are taking a bit step further that we are maintaining a server which provides all the information about the parking slots in a particular area

## 2. PROBLEM DEFINITION

Few years back there was no problem of the parking at all, because at that time there is no space problem so that no restriction for parking and only few vehicles so that no more load to take. When the population increase hence the production of vehicles (car's) also increased that's why the load on the road were increased, diminishing parking space and a myriad of confusing rules and regulations car owners find's it extremely difficult to find a solution to problem of parking.

## 3. PROBLEM ARCHITECTURE

The proposed architecture that we have conceived around n-Tier is given by:-

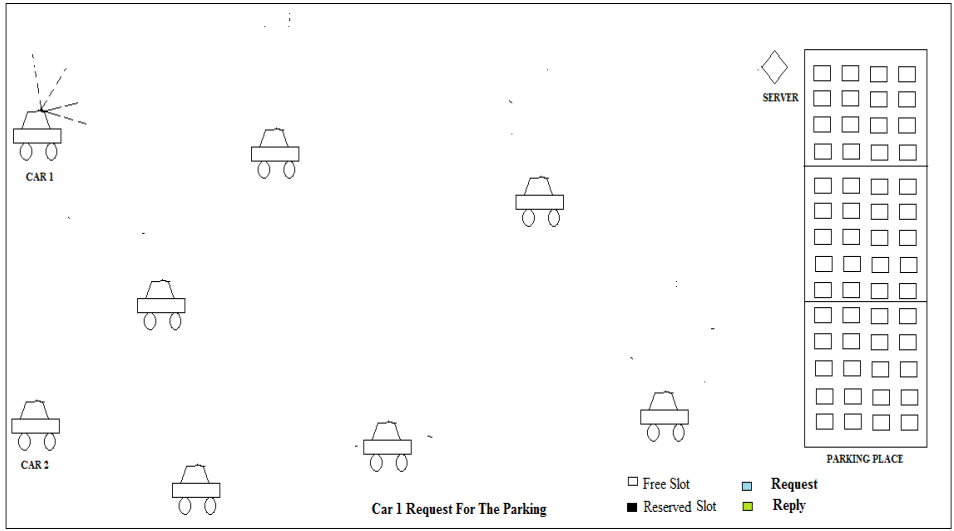

Figure 1:- Car Request For The Parking

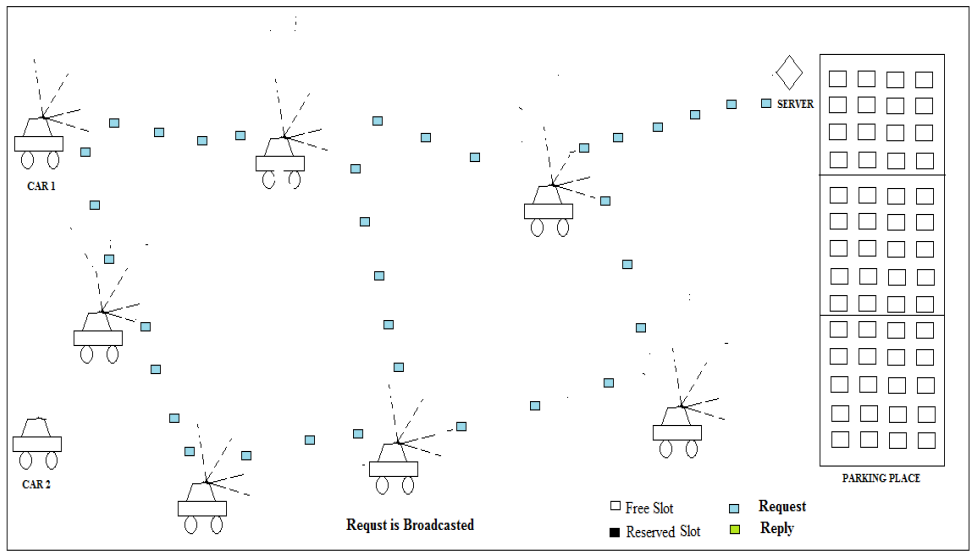

Figure 2: The Request Broadcasted through the Mobile vehicle's

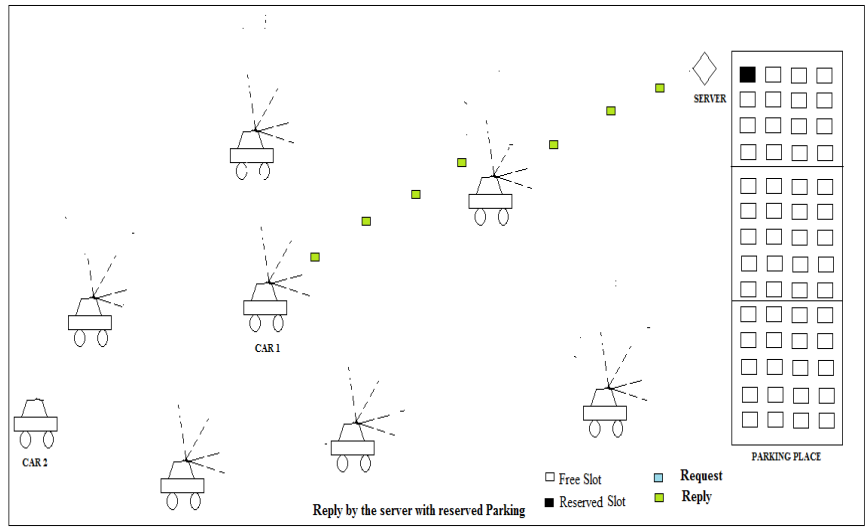

Figure 3: The Parking Server Replies For The Request

## 3.1 DESIGN DIAGRAM OF PROPOSED ARCHITECTURE

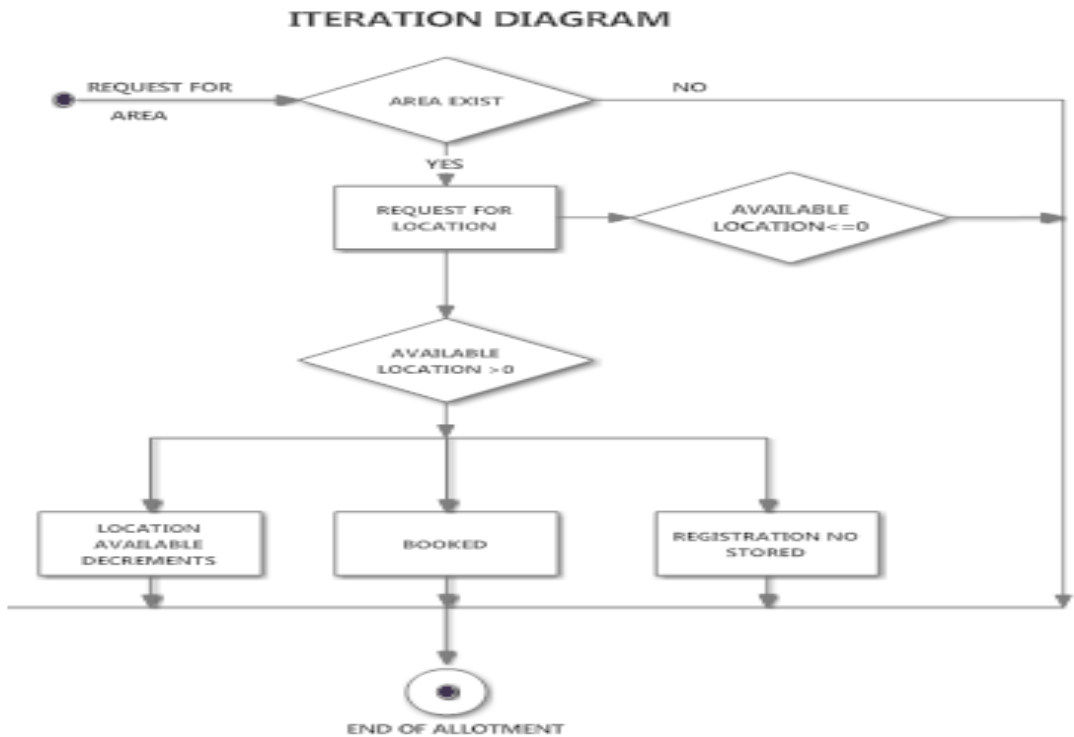

Figure 4:Iteration Diagram of Architecture

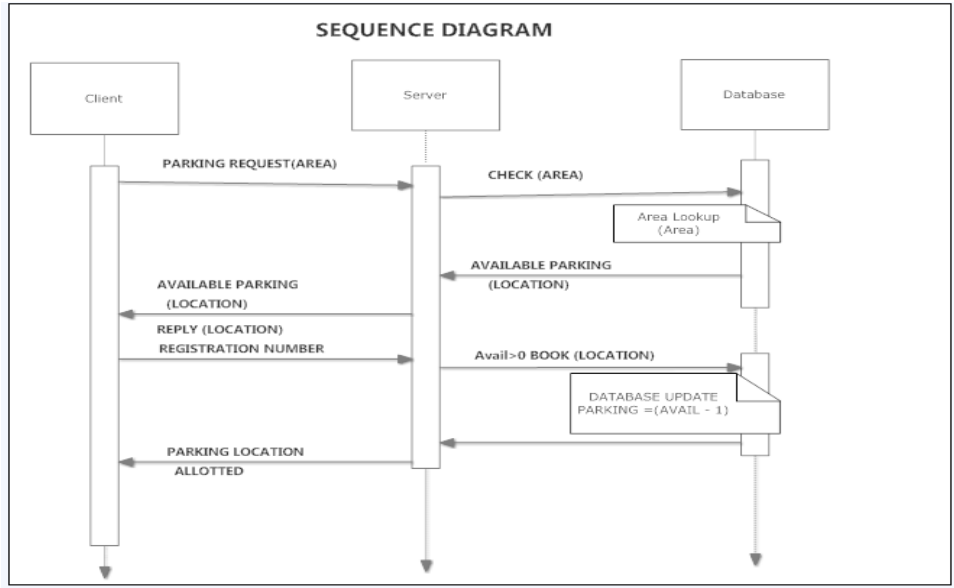

Figure 5: Sequence Diagram

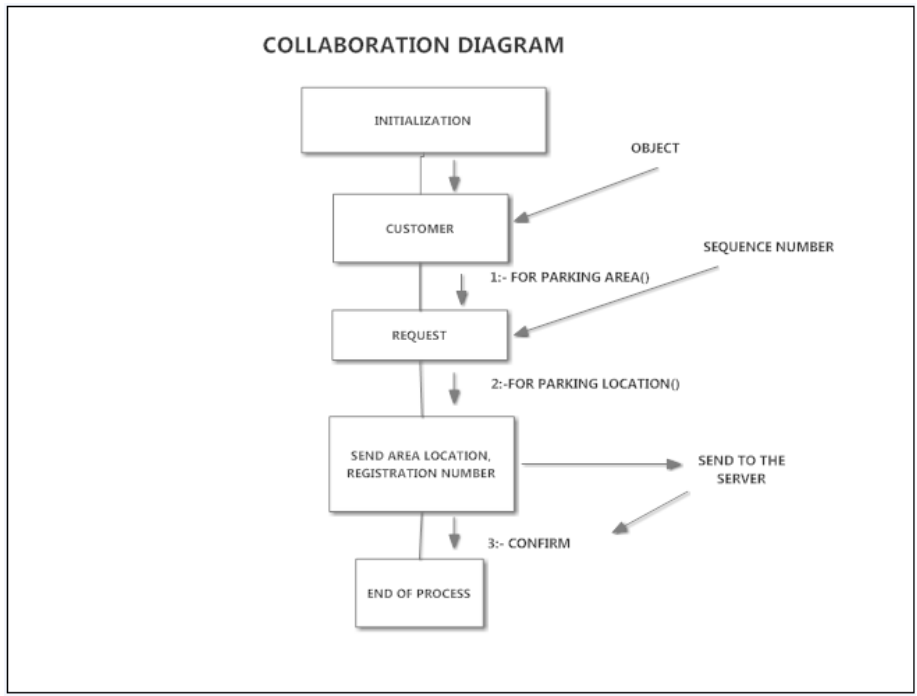

Figure 6: Collaboration Diagram

## 4. ACO ROUTING WITH ANYCAST SEMANTICS DELIVERY MODELS

For applying the membership model efficiently in our parking problem, anycast semantics is used to maintain the groups. DTNs has long delay for delivering the packets, at the time of delivery of packet there may be the member of the group change their groups. For the above its exigency to narrate a new anycast semantics model for DTN.

Here in fig 1 we can see that the request packets are being broadcasted through the mobile vehicles. These mobile vehicles are being divided into groups for forwarding the message. After making groups we apply a delivery model for delivering the request to the server by these mobile vehicles. In fig 2 if the packet has been broadcasted by the mobile vehicle, there is no need to apply the delivery models [8] on mobile vehicles but once the path is established, we apply the delivery models because the vehicles are mobile. At that time we need to apply a delivery models on that. It is described that the intended receiver should be clearly defined for a message as group membership changes when nodes join and leave the group. Hence, here three anycast semantics model are demonstrated that's permits senders to explicitly specify to intended receivers of the message.

### 4.1 Current membership model:

In this model the receiver of the packet should be destination group member at the time of message delivery. i.e When the packet reached to the intended receiver group the that particular mobile vehicle must be there in that group otherwise packet will be discarded.

### 4.2. Temporal Interval membership model:

In this model the sender sends a packet and also specify the time period during which the receiver must be the part of intended group at least once. If the receiver not a part of the intended group during that period of time defined by the sender the packet was discarded. i.e here the packet the time period is define for the delivery of packet. If the intended receiver atleast once between the time period define for receiving that packet hence packet will be discarded.

### 4.3. Temporal point membership model:

Its intended receiver at least should be a member of destination group at some time during membership interval. i.e in this model the packet will be forwarded to some of the mobile node of that intended group. So that the packet we be delivered to the intended receiver when it will join the group

By the use of these models we can deliver the message to the intended receiver correctly.

## 5. FUTURE WORK AND CONCLUSION

In future we would apply ACO to develop an efficient routing algorithm in addition to the algorithm stated in this paper. Thereafter using the concepts of Java Programming language, we would try to built a Open Source APL (Automated Parking Tool) around ACO technique to solve the realistic problem of parking of vehicles in parking/ bay areas. We would also try to implement the delivery models for delivering message to the appropriate vehicle and would design a dedicated server for each area which will contain all the information of that particular area. Messages will behave as an Ant and server as a food. A request would be generated and it will be broadcasted through the mobile vehicles for searching food. Once the request (ant) reaches the server (food) ,the reply will be unicasted and these paths are maintained by that particular group (nest) where the intended vehicle is situated .

## ACKNOWLEDGEMENTS

I honestly thank to all my faculty members of institution for their extra effort to make our session on line inspire of all ideas and the administration of GGSIPU University for providing academic environment to pursue our research activities.

## AUTHORS

**Rahul K Dixit**, Research Scholar, USICT, GGSIP University, Delhi-110078., AMIETE from I.E.T.E,Lodhi Road, Delhi Gate Scholar in 2011 and 2012.

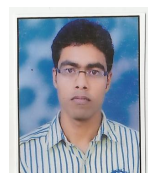

**Rahul Johari**, is working as a Assistant Professor in USICT, GGSIPU